\documentclass{PoS}
\usepackage{epsfig}

\newcommand{\intc}[1]{{\int\frac{d#1}{2i\pi}}}
\newcommand{\g}{\gamma}
\newcommand\lr[1]{{\left({#1}\right)}}

\title{Jet gap jet events at Tevatron and LHC}

\ShortTitle{Jet gap jet events at Tevatron and LHC}

\author{\speaker{Christophe Royon}\\
        CEA/IRFU/Service de physique des particules, CEA/Saclay,  91191 Gif-sur-Yvette cedex, France\\
        E-mail: \email{royon@hep.saclay.cea.fr}}

\abstract{We investigate diffractive events in hadron-hadron collisions, in which 
two jets are produced and separated by a large
rapidity gap.  
Using a renormalisation-group improved NLL kernel implemented in the HERWIG
Monte Carlo program, we show that the BFKL 
predictions are in good agreement with the Tevatron data, and 
present predictions which could be tested at the LHC.

}

\FullConference{The 2009 Europhysics Conference on High Energy Physics,\\
		 July 16 - 22 2009\\
		 Krakow, Poland}

\begin{document}

In a hadron-hadron collision, a jet-gap-jet event features a large rapidity 
gap with a high$-E_T$ jet on each side
($E_T\!\gg\!\Lambda_{QCD}$). Across the gap, the object exchanged in the 
$t\!-\!channel$ is color singlet and carries a large momentum transfer, 
and when the rapidity gap is sufficiently large the natural candidate in 
perturbative QCD is the
Balitsky-Fadin-Kuraev-Lipatov (BFKL) Pomeron \cite{bfkl}. Of course the total 
energy of the collision $\sqrt{s}$
should be big ($\sqrt{s}\gg E_T$) in order to get jets and a large rapidity 
gap.

Following the success of the forward jet and Mueller Navelet jet BFKL NLL
studies~\cite{nllfjus}, we use the implementation of the BFKL NLL kernel inside the
HERWIG~\cite{herwig} Monte Carlo to compute the jet gap jet cross section,
compare our results with the Tevatron measurement and make predictions at the
LHC~\cite{us}.

\section{BFKL NLL formalism}

The production cross section of two jets with a gap in rapidity between them reads
\begin{equation}
\frac{d \sigma^{pp\to XJJY}}{dx_1 dx_2 dE_T^2} = {\cal S}f_{eff}(x_1,E_T^2)f_{eff}(x_2,E_T^2)
\frac{d \sigma^{gg\rightarrow gg}}{dE_T^2},
\label{jgj}\end{equation}
where $\sqrt{s}$ is the total energy of the collision,
$E_T$ the transverse momentum of the two jets, $x_1$ and $x_2$ their longitudinal
fraction of momentum with respect to the incident hadrons, $S$ the survival probability,
and $f$ the effective parton density functions~\cite{us}. The rapidity gap
between the two jets is $\Delta\eta\!=\!\ln(x_1x_2s/p_T^2).$ 

The cross section is given by
\begin{equation}
\frac{d \sigma^{gg\rightarrow gg}}{dE_T^2}=\frac{1}{16\pi}\left|A(\Delta\eta,E_T^2)\right|^2
\end{equation}
in terms of the $gg\to gg$ scattering amplitude $A(\Delta\eta,p_T^2).$ 

In the following, we consider the high energy limit in which the rapidity gap $\Delta\eta$ is assumed to be very large.
The BFKL framework allows to compute the $gg\to gg$ amplitude in this regime, and the result is 
known up to NLL accuracy
\begin{equation}
A(\Delta\eta,E_T^2)=\frac{16N_c\pi\alpha_s^2}{C_FE_T^2}\sum_{p=-\infty}^\infty\intc{\g}
\frac{[p^2-(\g-1/2)^2]\exp\left\{\bar\alpha(E_T^2)\chi_{eff}[2p,\g,\bar\alpha(E_T^2)] \Delta \eta\right\}}
{[(\g-1/2)^2-(p-1/2)^2][(\g-1/2)^2-(p+1/2)^2]} 
\label{jgjnll}\end{equation}
with the complex integral running along the imaginary axis from $1/2\!-\!i\infty$ 
to $1/2\!+\!i\infty,$ and with only even conformal spins contributing to the sum, and 
$\bar{\alpha}=\alpha_S N_C/\pi$ the running coupling.

Let us give some more details on formula \ref{jgjnll}. The NLL-BFKL effects are 
phenomenologically taken into account by the effective kernels $\chi_{eff}(p,\g,\bar\alpha)$.
The NLL 
kernels obey a {\it consistency condition} which allows to reformulate the 
problem in terms of $\chi_{eff}(\g,\bar\alpha).$ The effective kernel
$\chi_{eff}(\g,\bar\alpha)$ is obtained from the NLL kernel $\chi_{NLL}\lr{\g,\omega}$ by 
solving the implicit equation
$\chi_{eff}=\chi_{NLL}\lr{\g,\bar\alpha\ \chi_{eff}}$ as a solution of the consistency condition.

In this study, we performed a parametrised distribution of $d \sigma^{gg\rightarrow gg}/dE_T^2$
so that it can be easily implemented in the Herwig Monte Carlo since performing the integral over
$\gamma$ in particular would be too much time consuming in a Monte Carlo. The implementation of the
BFKL cross section in a Monte Carlo is absolutely necessary to make a direct comparison with data.
Namely, the measurements are sensititive to the jet size (for instance, experimentally the gap size
is different from the rapidity interval between the jets which is not the case by definition in the
analytic calculation).

\begin{figure}
\begin{center}
\epsfig{file=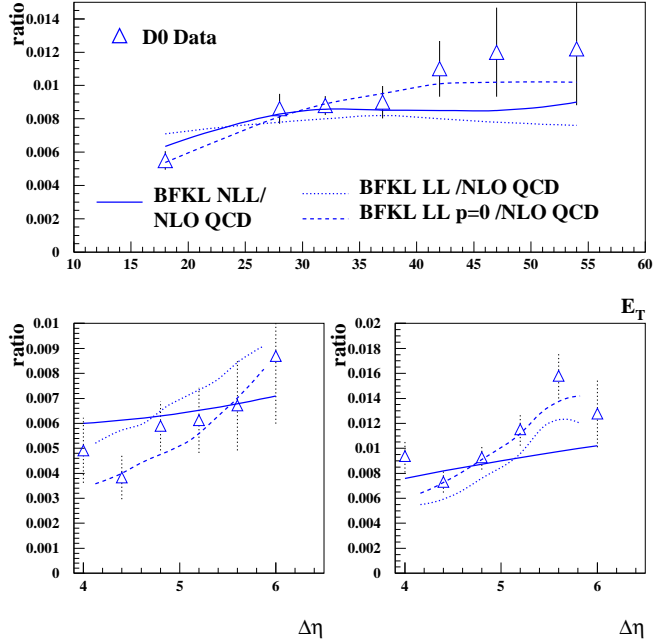,width=9.5cm}
\caption{Comparisons between the D0 measurements of the jet-gap-jet event ratio with the 
NLL- and LL-BFKL calculations. For reference, the comparison with the LL BFKL with only the
conformal spin component $p=0$ is also given.}
\label{fig1}
\end{center}
\end{figure}

\section{Comparison with D0 and CDF measurements}
Let us first notice that the sum over all conformal spins is absolutely necessary. Considering
only $p=0$ in the sum of Equation~\ref{jgjnll} leads to a wrong normalisation and a wrong jet $E_T$
dependence, and the effect is more pronounced as $\Delta \eta$ diminishes.

The D0 collaboration measured the jet gap jet cross section ratio with respect to the total dijet
cross section, requesting for a gap between -1 and 1 in rapidity, as a function of the second
leading jet $E_T$,
and $\Delta \eta$ between the two leading jets for two different low and high $E_T$ samples
(15$<E_T<$20 GeV and $E_T>$30 GeV). To compare with theory, we compute the following quantity
\begin{eqnarray}
Ratio = \frac{BFKL~ NLL~HERWIG}{Dijet~Herwig} \times \frac{LO~QCD}{NLO~QCD} 
\end{eqnarray}
in order to take into account the NLO order corrections on the dijet cross
sections, where $BFKL~ NLL$ $HERWIG$ and $Dijet~Herwig$ denote the BFKL NLL and the dijet cross section
implemented in HERWIG. The NLO QCD cross section was computed using the NLOJet++ program~\cite{nlojet}.

The comparison with D0 data~\cite{d0jgj} is shown in Fig.~\ref{fig1}. We find a good agreement between the data
and the BFKL calculation. It is worth noticing that the BFKL NLL calculation leads to a better result
than the BFKL LL one (note that the best description of data is given by the BFKL LL formalism 
for $p=0$ but it does not make sense theoretically to neglect the higher spin components and this
comparison is only made to compare with previous LL BFKL calculations).

\begin{figure}
\begin{center}
\epsfig{file=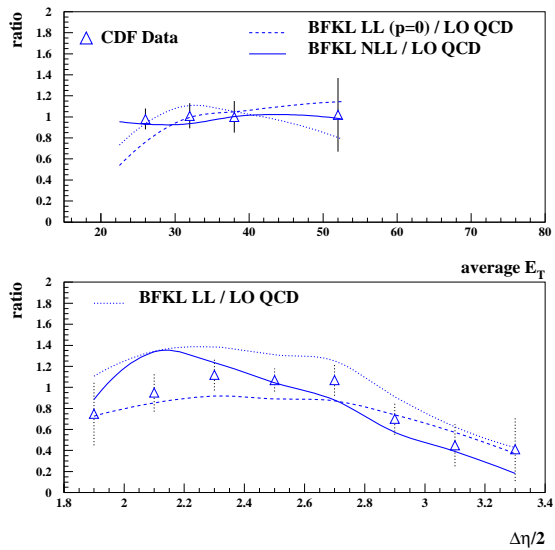,width=8.cm}
\caption{Comparisons between the CDF measurements of the jet-gap-jet event ratio with the 
NLL- and LL-BFKL calculations. For reference, the comparison with the LL BFKL with only the
conformal spin component $p=0$ is also given.}
\label{fig2}
\end{center}
\end{figure}

The comparison with the CDF data~\cite{d0jgj} as a function of the average jet $E_T$ and the
difference in rapidity between the two jets is shown in Fig.~\ref{fig2}, and the conclusion remains the same:
the BFKL NLL formalism leads to a better description than the BFKL LL one.

\section{Predictions for the LHC}
Using the same formalism, and assuming a survival probability of 0.03 at the LHC, it is possible to
predict the jet gap jet cross section at the LHC. While both LL and NLL BFKL formalisms lead to a
weak jet $E_T$ or $\Delta \eta$ dependence, the normalisation if found to be quite difference
leading to higher cross section for the BFKL NLL formalism.

\end{document}